\documentstyle[11pt,newpasp,twoside,epsf]{article}
\newcommand{\eg}{e.g.\thinspace}

\newcommand{\etal}{et al.\thinspace}

\begin{document}

\title{The Cosmological Origin of Disk Galaxy Scaling Laws}
\author{Matthias Steinmetz}
\affil{Steward Observatory, University of Arizona, 933 N Cherry Ave, Tucson, AZ
85721 USA}
\author{Julio Navarro}
\affil{Department of Physics and Astronomy, University of
Victoria, Victoria, BC V8P 1A1, Canada}

\begin{abstract}

We discuss possible origins of scaling laws relating structural properties of
disk galaxies within the context of hierarchically clustering theories of galaxy
formation. Using gasdynamical simulations that incorporate the effects of star
formation we illustrate these global trends and highlight the difficulties faced
by models that envision disk galaxies as the final outcome of a hierarchical
sequence of merger events. In particular, we focus on the cosmological origin of
the Tully--Fisher relation, and argue that this correlation between the total
luminosity and rotation speed of disk galaxies is a natural result of the
approximately scale free formation process of the massive halos they inhabit.
Although the slope and scatter of the Tully--Fisher relation can be readily
reproduced in hierarchical formation scenarios, the observed zero-point of the
relation is inconsistent with simulations of galaxy formation in Cold Dark
Matter universes, a difficulty that can be traced to the high central mass
concentration of dark halos formed in this cosmogony. This result indicates that
substantial revision to the new ``standard'' model of structure formation
($\Lambda$CDM) may be needed in order to accommodate observations on the scale
of individual disk galaxies.

\end{abstract}

\section{Introduction}
It is well known that the dynamical and structural properties of galaxies obey a
number of well defined scaling laws. Examples of these correlations include (i)
the Tully-Fisher (henceforth TF) relation relating the luminosity and rotation
speed of disk galaxies, (ii) the Fundamental Plane linking the surface
brightness, velocity dispersion, and size of elliptical galaxies, and (iii)
Schmidt's law, which relates gas surface density and star formation rates of
spiral galaxies. While some of these scaling laws can be deduced from fairly
basic physical principles, such as the virial theorem, in general the details of
their origin are only poorly understood.

In this contribution we report on recent progress identifying a cosmological
origin of the TF relation through high resolution numerical experiments.  The
present paper complements studies based on analytical and semianalytical models
reported elsewhere in this volume (see, \eg, contributions by F.~van den Bosch
and by H.~Mo) and improves upon earlier work on this topic by, e.g., Evrard
\etal (1994), Tissera \etal (1997), Elizondo \etal (1999), and Steinmetz \&
Navarro (1999). In \S2 we summarize the main features of our simulation
techniques and describe the results of our simulations while \S3 discusses our
findings. Section 4 summarizes our main results and conclusions.

\section{Numerical Simulations}

\subsection{The Numerical Method}

The simulations were performed using GRAPESPH, a code that combines the Smoothed
Particle Hydrodynamics (SPH) approach to numerical hydrodynamics with a direct
summation N-body integrator optimized for the special-purpose hardware GRAPE
(Steinmetz 1996). GRAPESPH is fully Lagrangian and highly adaptive in space and
time through the use of individual particle smoothing lengths and timesteps. It
is thus optimally suited to study the formation of highly non-linear systems
such as individual galaxy systems in a cosmological context.  The code used for
the simulations described in this paper include the self-gravity of gas, stars,
and dark matter, a full $3D$ hydrodynamical treatment of the gas, radiative and
Compton cooling, and a simple recipe for transforming gas into stars and for
incorporating the feedback of mass and energy into the gaseous component driven
by evolving stars.

\subsection{Star Formation and Feedback Recipes}

Star formation is modeled by creating new collisionless ``star'' particles in
Jeans-unstable, collapsing regions at a rate given by
$\dot{\varrho}_{\star}=c_{\star} \varrho_{\rm gas}/\max(\tau_{\rm
cool},\tau_{\rm dyn})$. Here $\varrho_{\rm gas}$ is the local gas density,
$\tau_{\rm cool}$ and $\tau_{\rm dyn}$ are the local cooling and dynamical
timescales, respectively, and $c_{\star}$ is a star formation ``efficiency''
parameter. After formation, ``star'' particles are only affected by
gravitational forces, but they devolve energy to their surroundings in a crude
attempt to mimic the energetic feedback from supernovae: $10^{49}$ ergs (per
$M_{\odot}$ of stars formed) are injected into their surrounding gas about
$10^7$ yrs after their formation. This energy is invested mostly in raising the
internal energy (temperature) of the gas, but a fraction $f_v$ is invested in
modifying the bulk motion of the gas surrounding star forming regions. For
details, see Steinmetz \& Navarro (1999) and references therein.

The star formation and feedback recipe described above involves two free
parameters: the star formation parameter $c_{\star}$ and the feedback parameter
$f_v$. We have performed three sets of simulations with various star formation
and feedback parameters; (i) $f_v=0$, $c_{\star}=0.05$, (ii) $f_v=0.2$,
$c_{\star}=0.05$, and (iii) $f_v=0.2$, $c_{\star}=1.00$. Each set includes about
35 galaxies with circular velocities between $80$ and $400$ km s$^{-1}$. Each of
these galaxies has, at $z=0$, at least 3000 star particles.
For these three parameter choices, Kennicutt's (1998) relation between the HI surface density
and the star formation rate per unit area (Schmidt's law) is roughly reproduced
in isolated galaxy test cases; however, the gas fractions of $z=0$ galaxy models vary
substantially (by more than a factor of 3) for the three different parameter
choices. 

\subsection{The Cosmological Model}

The cosmological models we investigate are two flavors of the Cold Dark Matter
(CDM) scenario; the traditional standard CDM model (``sCDM'', characterized by
$\Omega=1$, $h=0.5$, $\Omega_{\rm b}=0.0125 \, h^{-2}$, and $\Lambda=0$) and a
flat CDM model with an non-zero cosmological constant ($\Lambda$CDM; with
parameters $\Omega=0.3$, $h=0.7$, $\Omega_{\rm b}=0.0125 \, h^{-2}$, and
$\Lambda=0.7$). Both models are normalized to reproduce the observed number
density of massive galaxy clusters at $z=0$.

For each cosmological model and for each set of star formation and feedback
parameters, we identify $\approx 35$ dark matter halos and measure the circular
velocity of the central disks at a fiducial radius $r_{\rm gal}=15 \,
(V_{200}/220$ km s$^{-1}) \, h^{-1}$ kpc, where $V_{200}$ is the circular
velocity of the system at the ``virial radius'' where the mean inner overdensity
is 200 times the critical density for closure. The radius $r_{\rm gal}$ contains
most of the baryonic component of the luminous galaxy and is much larger than
the numerical softening; galaxy properties computed at $r_{\rm gal}$ are rather
insensitive to numerical limitations. The circular velocity profiles of
simulated galaxies are approximately flat near the center; indeed, the circular
velocity at $r_{\rm gal}=15 \, (V_{200}/220$ km s$^{-1}) \, h^{-1}$ kpc and at
$3.5 \, (V_{200}/220$ km s$^{-1}) \, h^{-1}$ kpc (corresponding to about 2 disk
scale lengths, the radius at which observed rotation velocities are typically
measured) differ by less than $20\%$ in all the systems we consider.

\subsection{A numerical Tully--Fisher relation}

Figure 1 (left panel) compares observational data with the simulated TF
relations obtained for two different feedback prescriptions in the sCDM
model. Corresponding data for the $\Lambda$CDM scenario are shown in figure 2
(left panel, open circles correspond to $c_*=0.05$ and $f_v=0$). This comparison
illustrates a number of interesting features.

\begin{itemize}
\item The slope of the numerical TF relation is in good agreement with the observed TF
relation, independent of the cosmological model and of the star formation and
feedback description. Only for very low rotation speeds ($V_{rot} \la 100\,$km
s$^{-1}$) a somewhat shallower slope can be observed in the case of a kinetic
feedback model. Also in agreement with observations, the numerical TF relations
in other bands (not shown) show a systematic steepening from the blue to the red
band passes (Steinmetz \& Navarro, 1999).
\item The scatter of the numerical TF relation is quite small. The rms scatter
in the I-band is only $0.25$ mag, even smaller than the observed scatter of
$\sim 0.4$ mag. This must be so if our results are to agree with observations:
scatter in the models likely reflects the intrinsic dispersion in the TF
relation, whereas the observed scatter includes the additional contribution from
observational errors. The small scatter is somewhat surprising, since the
fraction of baryons that cool and settle into the disk varies substantially
between individual halos. 
\item There is a serious discrepancy in the zero point of the TF relation. At a
given circular velocity, simulated galaxies are about 2 magnitudes (in $I$) too
dim.  The discrepancy is rather insensitive to the cosmological model and/or
adopted star formation and feedback recipe.
\end{itemize}

\begin{figure}
\plottwo{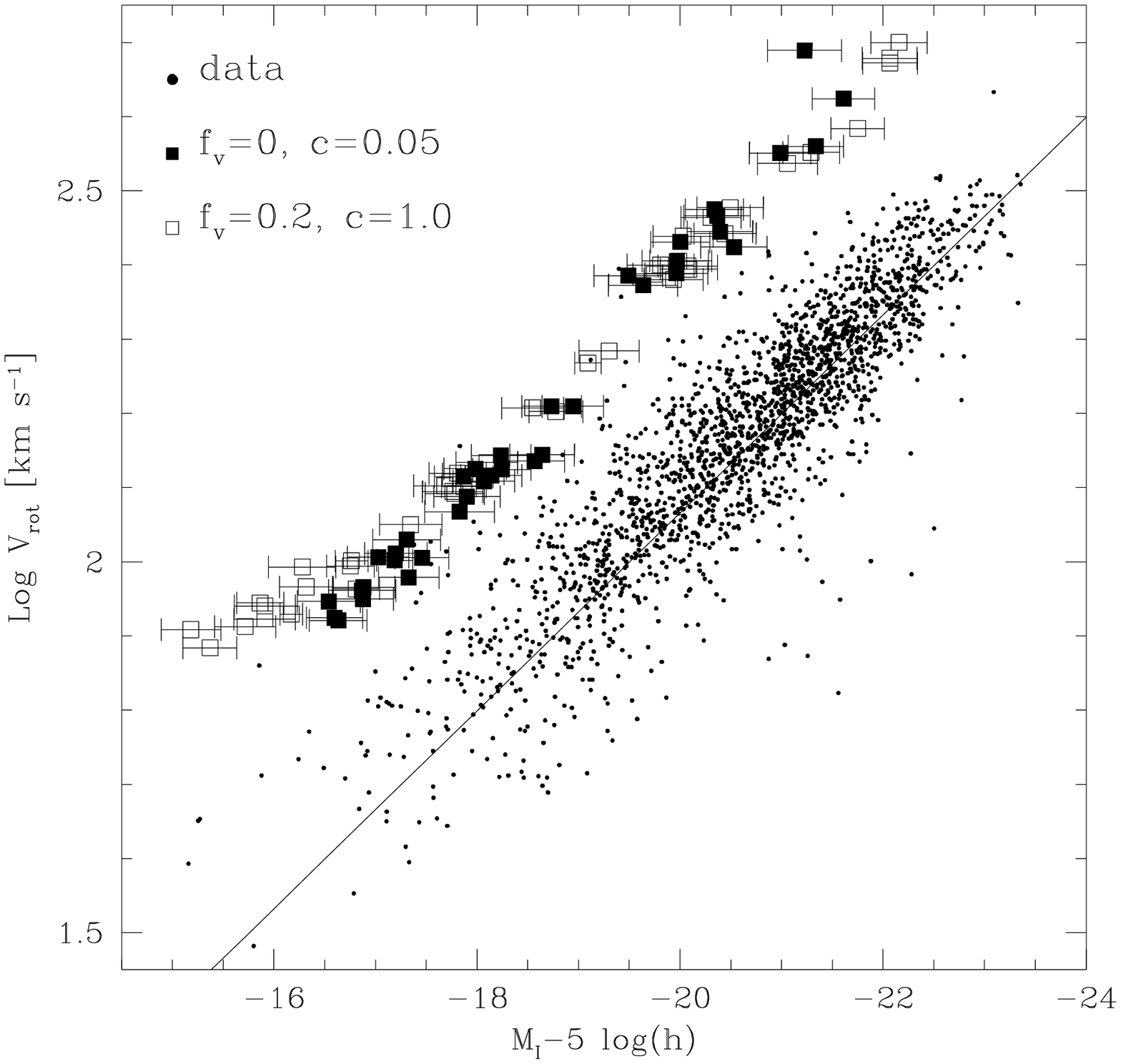}{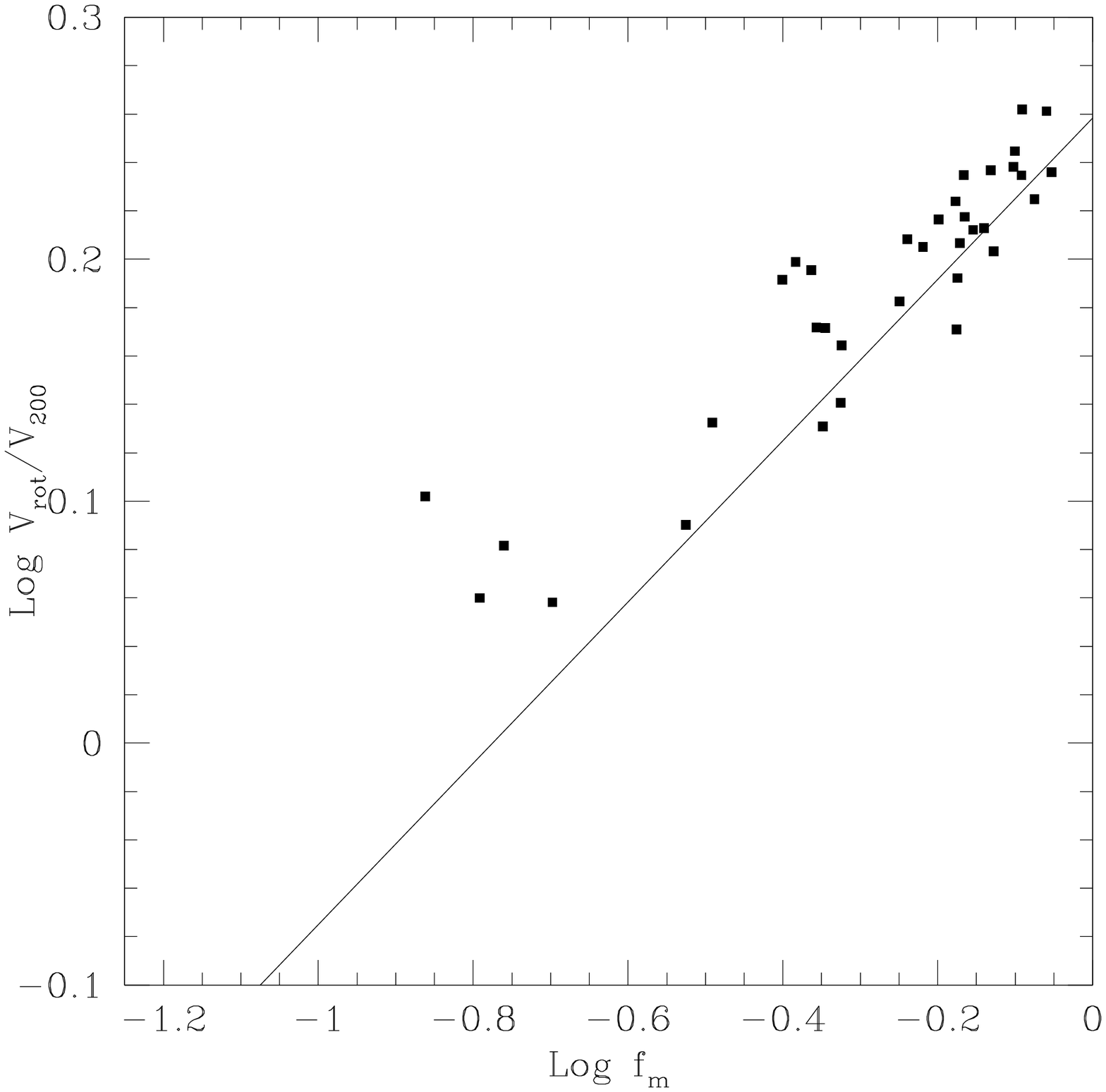}
\caption{Left: I band Tully--Fisher relation at $z=0$ for an
$\Omega=1,\Lambda=0$ CDM scenario. Error bars in the simulated dataset span the
difference in magnitude that results from adopting a Salpeter or a Scalo
IMF. Right: ratio of the circular velocities at $r_{\rm gal}$ and at the virial
radius (see text for definitions) versus the fraction of baryons that have
cooled, settled into the central disk and turned into stars. The solid line
corresponds to $V_{rot}/V_{200} \propto f_m^{1/3}$.}
\end{figure}

\section{TF Relation: Observations vs Numerical Experiments}

\begin{figure}
\plottwo{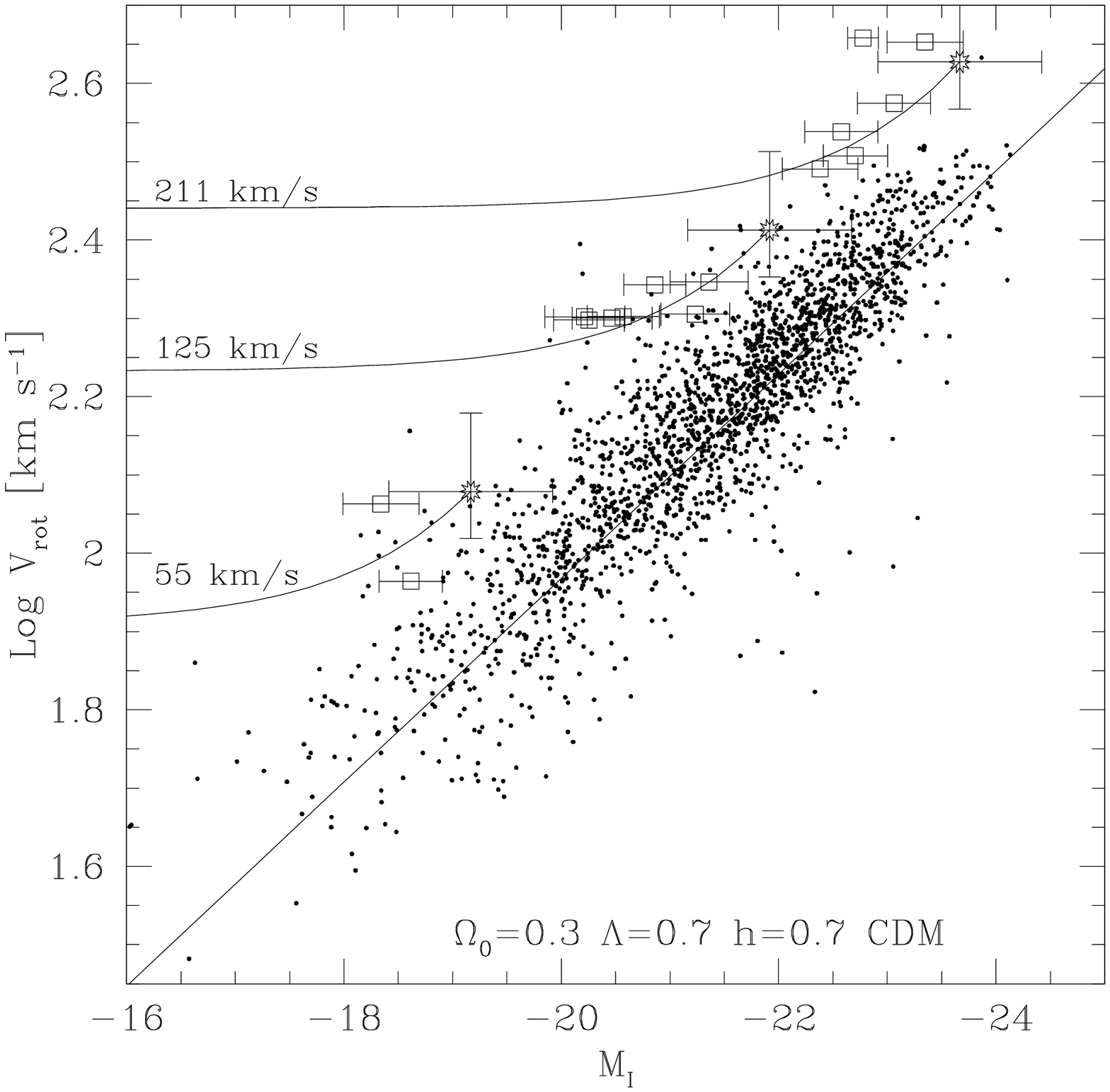}{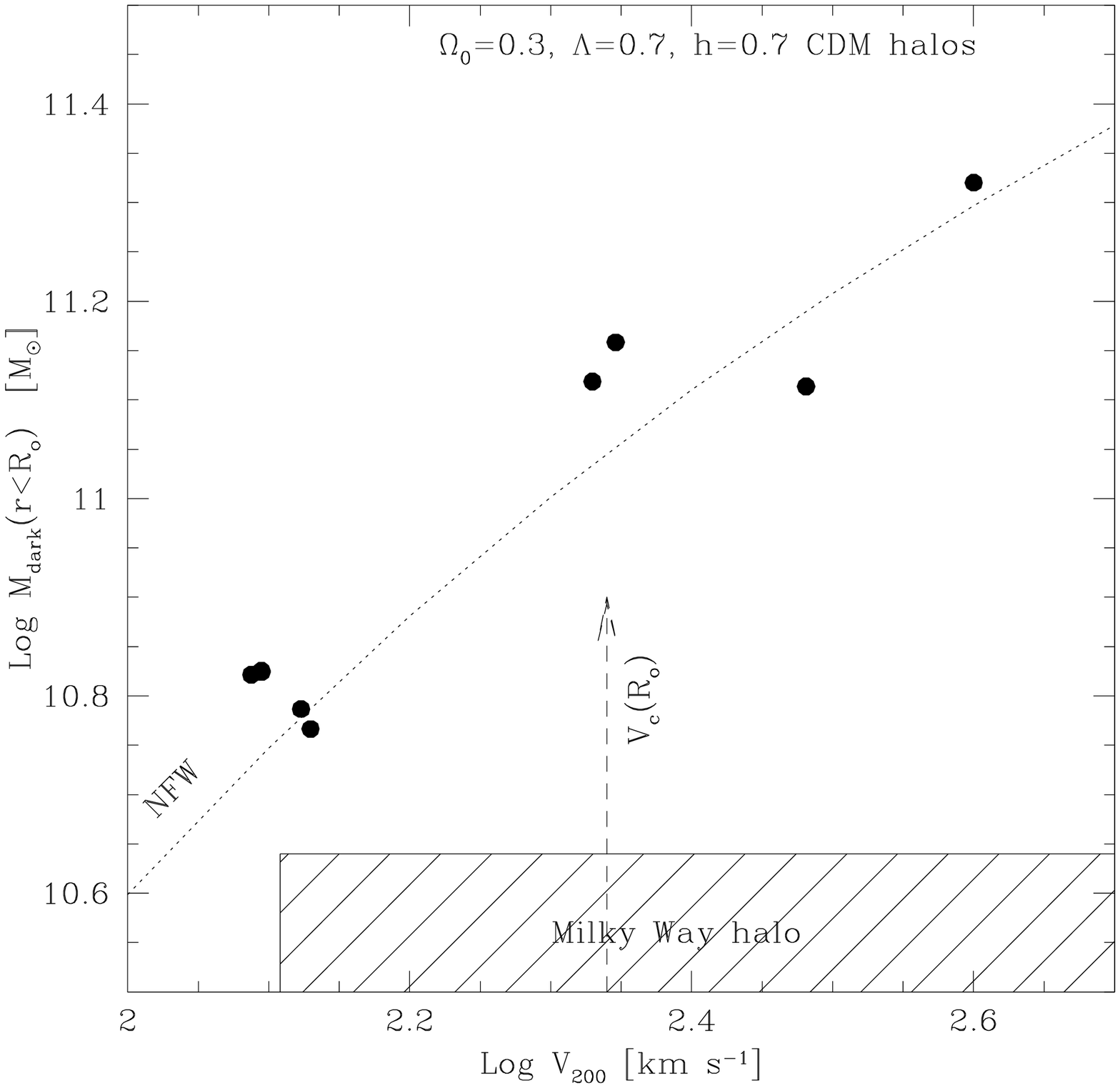}
\caption{Left: Right: I-band TF relation at $z=0$ for 
$\Omega=0.3,\Lambda=0.7$. See text for details.  Right: Dark mass enclosed
within a radius $R_{o}= 8.5$ kpc, the Sun's distance from the center of the
Milky Way, versus the circular velocities of $\Lambda$CDM halos. The shaded
region highlights the allowed parameters of the dark halo surrounding the Milky
Way.  The filled circles show the loci of $\Lambda$CDM halos as determined from
high-resolution N-body simulations. The dotted line is the circular velocity
dependence of the dark mass expected inside $R_{o}$ for halos that follow the
density profile proposed by NFW.}
\end{figure}

In order to further our insight into the meaning of the successes and failures
of out TF modeling we shall focus on the following questions. (i) Why is the
simulated TF slope in such good agreement with observations? (ii) Why is the
scatter in the numerical TF relation so small, in particular considering that
the fraction of baryons inside the virial radius effectively accreted into the
central galaxy, $f_m$, varies significantly from halo to halo? (iii) Why is the
zero-point of the numerical relation in disagreement with observations?

\subsection{The Slope}

The slope of the numerical TF relation can be understood as a direct consequence
of the cosmological equivalence between mass and circular velocity (see, e.g.,
Mo, Mao \& White 1998). This equivalence is a consequence of the finite age of
the universe, which imposes a maximum radius (approximately equal to the
``virial radius'') from where matter can accrete to form a galaxy (for an
alternative view in which the slope is derived as a result of self-regulated
star formation in disks of different mass, see Silk 1997).  At $z=0$, the
circular velocity at the virial radius ($V_{200}$) and its enclosed mass
($M_{200}$) are then equivalent measures of the system's mass related by
\begin{equation}
M_{200}= 2.33 \times 10^5 \biggl({V_{200} \over {\rm km \, s}^{-1}}\biggr)^3\,
h^{-1} M_{\odot}.
\end{equation}
Given this scaling, if disk rotation speeds are proportional to the circular
velocity of their surrounding halos ($V_{rot} \propto V_{200}$) and disk
(stellar) masses are proportional to the mass of the halos, $M_{disk} \propto
M_{200}$, then one expects $L_I=\Upsilon_I^{-1} M_{disk} \propto V_{\rm rot}^3$,
under the assumption that the I-band stellar mass-to-light ratio, $\Upsilon_I$,
does not vary dramatically with galaxy luminosity. Such velocity dependence is
similar to that of the observed TF relation, which can be approximated by
\begin{equation}
L_I \approx 2.4 \times 10^{10} \left({V_{\rm rot} \over 200 \, {\rm km \,
s}^{-1}}\right)^3 h^{-2} L_{\odot},
\end{equation}
(see solid lines in the left panels of Figures 1 and 2). Assuming that the
baryon density parameter is $\Omega_b \approx 0.0125 h^{-2}$, as suggested by
Big Bang nucleosynthesis studies of the abundance of the light elements
(Schramm \& Turner, 1998), the fraction of baryons transformed into stars in
disk galaxies is given by,
\begin{equation}
f_{m} \approx \Upsilon_I \, \Omega_0 \, h \, \left(\frac{V_{\rm
rot}}{V_{200}}\right)^3.
\end{equation}
Under the assumption that $V_{rot}=V_{200}$ and adopting $\Upsilon_I
\sim 2$ in solar units, as suggested by stellar population synthesis models,
eq.~3 implies that almost all the baryons in sCDM halos have been transformed
into stars, but that only about $40\%$ have suffered that fate in $\Lambda$CDM
halos (see Navarro \& Steinmetz, 2000a,b for details). The low efficiency in
the transformation of baryons into stars (especially in the case of $\Lambda$CDM
halos) should be contrasted with the relatively large specific angular momenta
of disks and halos of similar circular velocities. Although only $40\%$ of the
available baryonic mass is collected at the center of a $\Lambda$CDM halo,
analytical estimates indicate that those baryons are able to acquire as much as
$\sim 100\%$ of the specific angular momentum of the halo (Mo et al 1998). This
appears to indicate that disk galaxies are assembled preferentially from
high-angular momentum content gas, a feature that has proved rather difficult to
reproduce in numerical simulations (see, e.g., Navarro \& Steinmetz
1997). 

\subsection{The Scatter}

In the context of the model described in \S3.1, a number of possible sources of
scatter in the TF relation may be identified; variations in the ratio between
disk rotation speed and halo circular velocity, systematic variations in the
stellar mass-to-light ratio of galaxies with similar rotation speeds, and
disparities in the fraction of baryons $f_m$ that cool and settle into the disk.
The latter can be easily measured in our simulations, and is found to vary by
almost a factor of $2$ from galaxy to galaxy. Taken at face value, this would
imply a scatter in the TF relation of approximately $0.8$ mag, larger than the
observed value of $\sim 0.4$ mag. However, as discussed above, this is not the
case; the measured scatter of the numerical TF relation is actually
significantly smaller than the observed one.

The small scatter in the numerical TF relation introduced by variations in $f_m$
results from a tight relation between $f_m$ and the rotation velocity of the
galactic disk. The higher the fraction of baryons that assemble into the central
stellar disk the higher the luminosity of the galaxy, but also the higher its
circular velocity, due to the gravitational contribution of this additional
material. This effect is further amplified by the dark matter's response to this
extra gravitational pull.  This is illustrated in Figure 1 (right panel), which
shows the ratio of disk rotation speed (measured at $r_{\rm gal}$) and the
virial velocity, $V_{200}$, versus $f_m$. The velocity ratio scales
approximately as $f_m^{1/3}$ (shown as a solid line), indicating that variations
in the star to dark matter fraction result in shifts parallel to the TF
relation, adding very little scatter to the original relation. The scatter in
the numerical TF relation is thus largely the result of (small) differences in
the assembly and star formation history of the halos that these galaxies
inhabit.

\subsection{The Zero Point}

Although the slope and scatter of the numerical TF relation compare favorably
with observation, simulated galaxies are, at given rotation speed, almost two
magnitudes fainter than observed. This serious discrepancy seems to be
associated with the large concentration of dark matter near the center of Cold
Dark Matter halos. When these large amounts of dark matter are added to the
substantial mass in baryons needed to build a bright spiral the rotational
speeds that result are inconsistent with the observed TF relation.

This is shown in Figure 2 (left panel), where we show where hypothetical disk
galaxies built at the center of three representative dark halos of different
circular velocity would lie in the $M_I-V_{\rm rot}$ plane. The three solid
curves in this panel illustrate the loci of galaxies of different luminosity in
this plane, under the assumption of constant stellar mass-to-light ratio
($\Upsilon_I=2$ in solar units). Along each curve (from left to right) the
stellar mass of the disk varies from $\sim$ zero to the maximum value compatible
with the baryonic content of the halo, the rightmost point of each curve. As the
disk mass increases, each hypothetical galaxy moves from left to right across
the plot. When the disk mass becomes comparable to the dark mass inside the
optical radius of the galaxy the curve inches upwards and becomes essentially
parallel to the observed TF relation. Because of this, even under the extreme
assumption that galaxies contain {\it all} available baryons in each halo,
simulated disks are almost two magnitudes fainter than observed.

Increasing the baryonic mass of a halo further has virtually no effect on this
conclusion, since in this case model galaxies would just move further along
paths approximately parallel to the TF relation. As mentioned above, the main
reason for the discrepancy is the large central concentration of dark matter in
CDM halos.  This is shown in Figure 2 (left), where the circular velocity in the
inner few kpc of a $V_{200}=125$ km s$^{-1}$ halo contributed by the dark matter
{\it alone} is shown to be of order $170$ km s$^{-1}$ (see middle solid line in
this panel). Such large velocities only increase further as a result of the
assembly of the luminous component, leading to the large zero-point disagreement
observed in the left panels of Figures 1 and 2. Therefore, unless one (or more)
of the assumptions above is grossly in error, disk galaxies assembled at the
center of halos formed in the cosmological models we explored (sCDM and
$\Lambda$CDM) cannot match the observed Tully--Fisher relation.

Perhaps the most uncertain step in this argument is the stellar mass-to-light
ratio adopted for the analysis. The horizontal ``error bar'' shown on the
starred symbols in Figure 2 (left) illustrates the effect of varying the
$I$-band mass-to-light ratio by a factor of two from the fiducial value of $2$
in solar units. This is not enough to restore agreement with observations, which
would require $\Upsilon_I \sim 0.4$, a value much too low to be consistent with
standard population synthesis models. The vertical ``error bars'' illustrate the
effect of varying the ``concentration'' of each halo by a factor of two. Even
with this large variation in halo structure, the model disks fail to reproduce
the observations.

The only way to collect a massive disk galaxy without increasing $V_{\rm rot}$
significantly over $V_{200}$ is to have dark halos that are less centrally
concentrated than those formed in the sCDM or in the $\Lambda$CDM scenario. We
note that this problem is unlikely to be solved just by adjusting cosmological
parameters of the CDM model. The combination of parameters needed to reproduce
the present-day abundance of galaxy clusters is such that the central densities
of galaxy-sized dark halos is approximately independent of $\Omega_0$ and of the
value of $H_0$ (Navarro 1999). This explains why there is no noticeable
difference in the zero point of the sCDM and $\Lambda$CDM simulated TF relations
(Figures 1 and 2, respectively). The difficulties may be even more generic:
modifications to the CDM scenario that may potentially solve the problems
described here (\eg tilted power spectra, additional hot dark matter,
annihilating dark matter, to name a few) will delay the formation epoch of dark
matter halos to lower redshift. This may hinder the formation of massive
galaxies at high redshift, at odds with the mounting evidence that such galaxies
are fairly common at $z\ga 3$ (see, e.g., Steidel et al 1998).

\subsection{Application to the Milky Way halo}

The conclusion that cold dark matter halos are too centrally concentrated can be
independently checked by using observations of the dynamics of the Milky Way that
constrain the total dark mass within the solar circle ($R_{o}$). A direct
estimate can be made by assuming that the halo is spherically symmetric and by
combining the local density of the disk derived from ``Oort limit'' analysis
with the IAU-sanctioned values of $R_o=8.5$ kpc and $V_{\rm rot}(R_o)=220$ km
s$^{-1}$. This implies that the dark mass within $R_o$ cannot exceed
$M_{dark}(r<R_o)=5.2 \times 10^{10} M_{\odot}$ (Navarro \& Steinmetz
2000a). 

Figure 2 (right panel) compares this value with dark masses inside $8.5$ kpc
measured directly from N-body simulations of $\Lambda$CDM halos. The conclusion
is clear: halos with $V_{200} \approx 220$ km s$^{-1}$ have on average {\it three
times} more dark matter inside $R_o$ that allowed by observations. Only halos
with $V_{200} < 100$ km s$^{-1}$ are eligible as hosts of the Milky Way, but
they can be ruled out on the basis of the total (baryonic) mass of the luminous
component, $M_{disk}\approx 6 \times 10^{10} M_{\odot}$. This mass implies a
strict minimum halo mass through the universal baryon fraction of the universe,
which, expressed in terms of circular velocity, indicates that the halo of the
Milky Way must have $V_{200} \gg 130$ km s$^{-1}$. As illustrated in Figure 2
(right), these constraints on the halo of the Milky Way are inconsistent with
the results of N-body simulations.

\section{Summary and Conclusions}

We have studied the formation of disk galaxies in hierarchical clustering
scenarios using high resolution hydrodynamical simulations that include the
effects of star formation and supernova feedback. We report here on our
contribution to understanding the cosmological origin of disk galaxy scaling
laws and, in particular, of the Tully--Fisher relation. Our findings can be
summarized as follows.

\begin{itemize}

\item
Adopting star formation recipes that match observational constraints (\eg\
Kennicutt's law), the slope and scatter of the TF can be easily reproduced in
such numerical simulations. The slope of the I-band TF relation is essentially
determined by cosmological equivalence between the mass and the circular
velocity of dark matter halos. Variations in the star formation history as a
function of galaxy mass only introduces a slight modulation in this slope,
resulting in shallower slopes in the bluer pass bands.

\item
The tightness of the TF relation and its weak sensitivity to the star formation
and feedback prescription is largely the result of the kinematic response of the
dark matter halo to changes in the fraction of baryons that cool and settle in
the central disk. The increase in the rotation velocity of the galaxy due to the
self gravity of the disk and to the contraction of the dark matter halo scales
in such a way that, for relevant baryon fractions, leads to displacements
parallel to the TF relation, so that variations in the disk mass assembled
within each dark halo affect the slope of the relation only weakly and introduce
little additional scatter.

\item
Despite the success in reproducing its slope and scatter, galaxies formed inside
cold dark matter halos exhibit a substantial zero-point offset from the observed
TF relation. This is due to the large central concentrations of simulated dark
matter halos, which, we show, are also in disagreement with dynamical evidence
on the structure of the Milky Way's halo.

\end{itemize}

Although our conclusions are strictly valid for the two cosmological models we
have investigated (sCDM and $\Lambda$CDM) they are likely to extend to all Cold
Dark Matter models normalized to reproduce the observed abundance of massive
galaxy clusters. This implies that the discrepancies we highlight here signal
that the Cold Dark Matter paradigm, so successful at reproducing observations on
large scales, may actually be unable to accommodate the constraints placed by
detailed dynamical studies of individual galaxies. A thorough revision of the
validity and applicability of the Cold Dark Matter scenario on the scale of
individual galaxies seems warranted.

\end{document}